\documentstyle[12pt]{article}
\begin{document}
\normalsize
\begin{center}
{\Large \bf 
Towards a sufficient criterion for collapse in $3D$ Euler equations
}
\end{center}
\vspace{0.5cm}
\begin{center}
{\bf E.A. Kuznetsov\footnote{ e-mail: kuznetso@itp.ac.ru}}\\
{\it Landau Institute for Theoretical Physics} \\
{\it 2 Kosygin str., 117334 Moscow, Russia}
\end{center}


\begin{abstract}
A sufficient integral criterion for a blow-up solution of 
the Hopf equations (the Euler equations 
with zero pressure) is found. This criterion shows that
a certain positive integral quantity blows up in a finite time under specific
 initial conditions. Blow-up of this quantity means that
solution of the Hopf equation in $3D$ can not be continued in
the Sobolev space $H^2({\cal R}^3)$ for infinite time.
\end{abstract}

\section{Introduction}

In 1984 Beale, Kato and Maida  \cite{1} showed that sufficient and necessary
condition for a smooth solution  to 
the $3D$ Euler equations 
for ideal incompressible fluids on the  time interval $[0,t_0]$
is the finiteness of the integral,
\begin{equation}
\label{1}
\int\limits^{t_0}_0~\mbox {\rm
sup}_r|{\bf\omega}|~ dt <\infty .
\end{equation}
Here ${\bf\omega}$ is the vorticity, connected with the velocity field
${\bf v}$ by the standard relation:
\begin{displaymath}
{\bf\omega} = \mbox {\rm curl}~{\bf v}.
\end{displaymath}
If the integral (\ref{1}) is divergent, then the vorticity blows up (or
collapses) in a finite time.

Proof of this criterion is based on the local existence theorem \cite{2}.
According to this theorem a smooth solution to the 3D incompressible 
Euler equations exists if the initial conditions ${\bf v}(0)$
belong to the Sobolev space $H^q({\cal R}^3)$ for $q\geq 3$
and the solution itself is from the class,
$$
{\bf v}\in C([0,t_0];H^q)\cap C^1([0,t_0];H^{q-1}),
$$
where the norm of $H^q$
is defined as follows,
\begin{equation}
\label{norm}
\Vert {\bf v }\Vert_{H^q({\cal R}^3)}=
\Biggl ( \sum_{\alpha\leq q}\int~\biggl ( \nabla^{\alpha} {\bf v}\biggr )^2~d{\bf r}
\Biggr )^{1/2},
\end{equation}
with $\alpha$ being a multi-index.
In accordance with  \cite{1}, the violation of this property leads  to
divergence of the integral (\ref{1}) and to collapse for
vorticity.

However, in order to use the criterion (\ref{1}), effectively one
needs to have an explicit solution to the Euler equation, that is
practically very difficult or impossible. Therefore, the main task is 
to find { \it sufficient} conditions which guarantee blow-up of the vorticity. 
The first step in this direction would be to construct such criteria 
for models more simple than the
incompressible Euler equations which is one of the main aims of 
this  paper.

We consider the simplest variant  -- the Euler equations for a
compressible fluid without pressure (the hydrodynamics of dust) which 
are  sometimes called
the Hopf equations:
\begin{equation}
\label{2}
\frac{\partial\rho}{\partial t}+
\mbox {\rm div}~\rho{\bf v} =0,
\end{equation}
\begin{equation}
\label{3}
\frac{\partial\bf v}{\partial t}+
({\bf v}\cdot\nabla ) {\bf v}=0.
\end{equation}
These equations can be successfully solved by means of the Lagrangian
description. In terms of Lagrangian variables the equation (\ref{3})
describes the motion of free fluid particles.  For this reason 
breaking is possible in this model: it happens when 
trajectories of fluid particles  intersect. In terms of the 
mapping, describing the change of variables
from the Eulerian description to the Lagrangian one, 
the process of breaking corresponds to a vanishing Jacobian, $J$, of the mapping.  
This process can be considered as a
collapse in this system:  the
density $\rho$ as well as the velocity gradient become infinite at 
a finite time
in the points where the Jacobian vanishes. By virtue 
of this the breaking is sometimes  called 
gradient catastrophe. 

In this paper  we 
construct sufficient integral criteria for breaking 
within the model (\ref{2}),(\ref{3}) and  verify what the functional
conditions of the theorem \cite{1}
correspond to those for the given system. Our main conjecture 
is the following:  breaking in the 3D case within the system  
yields  a violation of
Sobolev space already for $q=2$ instead of $q\geq 3$ in the theorem \cite{1}. 

To complete the introduction we would like to note that 
the model (\ref{2}), (\ref{3}) has a lot of astrophysical applications. 
 According to the pioneering idea of Ya.B. Zeldovich \cite{4}, 
the formation of proto-galaxies due to breaking of stellar dust
can be described by the system (\ref{2}), (\ref{3}) (see also the review \cite{5}).
It should be noticed also that recently  \cite{6,ZKP}
the mechanism of breaking was used to explain collapse
in incompressible fluids when, instead of the breaking of fluid particles, 
the breaking of 
vortex lines can happen which results in a blow-up of vorticity.  
Therefore we believe that the results presented in this paper can
 also be useful
from the point of view of collapse in incompressible fluids.

\section{One-dimensional analysis}

We begin with a one-dimensional calculations  where Eq. (\ref{3}) 
has the form
\begin{equation}
\label{4}
 v_t+vv_x=0 ~.
\end{equation}
Consider the integrals 
\begin{displaymath}
I_n=\int v^n_x~dx
\end{displaymath}
with integer $n$, where integration is assumed to be  
from $-\infty$ to $+\infty$.

It is easy to show that the evolution of these integrals is defined by the
relations:
\begin{equation}
\label{5}
\frac{dI_n}{dt}=-(n-1)~I_{n+1},
\end{equation}
which can be obtained by means of (\ref{4}) and integration by parts.

The relations for $n=2,3$ read:
\begin{equation}
\label{6}
\frac{d}{dt}\int v^2_x~dx=-\int v^3_x~dx~,
\end{equation}
\begin{equation}
\label{7}
\frac{d}{dt}\int v^3_x~dx=-2\int v^4_x~dx ~.
\end{equation}

These differential relations allow one to find a closed differential
inequality for the integral $I_2=\int~v^2_x~dx$. 
Applying the Cauchy-Bunyakovsky inequality to (\ref{6}) gives the estimation
\begin{equation}
\label{8}
\frac{dI_2}{dt}\le \Biggl ( \int v^4_xdx \Biggr )^{1/2}I^{1/2}_2.
\end{equation}
Substituting, 
$$
I_4=\int v^4_x~dx=\frac{1}{2}
\frac{d^2I_2}{dt^2},
$$ 
into  (\ref{8}) we arrive at the closed differential
inequality for $I_2$:
\begin{equation}
\label{9}
I_2\cdot\frac{d^2I_2}{dt^2} - 2\Biggl ( \frac{dI_2}{dt} \Biggr
)^2\ge 0.
\end{equation}
This inequality is solved by means of change of variables
\begin{displaymath}
I_2=X^{\alpha}>0,
\end{displaymath}
with unknown exponent $\alpha$. We will define the exponent by
requiring the absence in (\ref{9}) of the terms $\sim X^{\alpha -2}X^2_t$.
Hence $\alpha =-1$ and the inequality becomes,
\begin{equation}
\label{10}
X_{tt}\le 0.
\end{equation}
Using a mechanical interpretation of 
$X$ as the coordinate of a particle and taking into account
that the particle
 acceleration is negative, we can immediately 
conclude that the particle can reach 
origin $X=0$ in a finite time if the initial particle velocity 
$X_t(0)$ is negative. For $I_2$ this  means
a blow-up:
\begin{displaymath}
I_2=\frac{1}{X} \to \infty.
\end{displaymath}
Hence, by elementary calculation, we can estimate the collapse
time $t_0$: 
\begin{displaymath}
t_0<\frac{X(0)}{|X_{t}(0)|}\equiv\frac{I_2(0)}{|I_{2t}(0)|}.
\end{displaymath}
After multiplying (\ref{10}) by $X_t$ $(<0)$ and integrating the result
over time,  we arrive at the estimation from above for 
$I_2$ :
\begin{equation}
\label{11}
I_2(t)\le\frac{I^2_2(0)}{| I_{2t}(0)| (t_0-t)}.
\end{equation}
Thus, the blow-up of $I_2$ takes place if the initial
velocity $X_t(0)$ is negative which is equivalent to the initial
condition,
\begin{displaymath}
I_{2t}(0)>0 
\end{displaymath}
or,
$$
I_3(0)=\int v^3_xdx<0.
$$
If the initial distribution of $v_0(x)$ is symmetric with respect to $x$, then
$I_3(0)\equiv 0$. However, as follows from (\ref{6}), $I_3$ becomes negative already at
$t=+0$ that results in a  blow-up of the integral $I_2$.

To complete this section, let us compare the estimate (\ref{11}) with the exact 
time dependence of
$I_2(t)$  near the breaking point.

In order to define this dependence, we differentiate  (\ref{4})
with respect to $x$ that results in the equation,
\begin{equation}
\label{12}
\frac{d v_x}{dt}=-v^2_x, \qquad
\Biggl ( \frac{d}{dt}=\frac{\partial}{\partial t}+
v\frac{\partial}{\partial x} \Biggr ).
\end{equation}
After integration we have 
\begin{equation}
\label{12'}
v_x=\frac{v'_0(a)}{1+v'_0(a)t},
\end{equation}
where $a$ is the initial coordinate of the fluid particle at $t=0$,
$v_0(a)$ is the initial velocity, the prime denotes a derivative.
The denominator in (\ref{12'}), $1+v_{a}(a)t$, represents 
the Jacobian $J$ of the mapping,
$$
x=a+v(a)t.
$$
The Jacobian  tends to zero for the first time
at $t=t_0$, defined by
\begin{equation}
\label{min}
t_0=\min_{a}[-v^{-1}_{a}(a)]>0.
\end{equation}
Near the singular point, $a=a_0$, corresponding to the minimum,
(\ref{min})  $J$ can be approximated by the expression,
$$
J\approx \alpha (t_0-t) + \gamma (a-a_0)^2.
$$
Here 
$$
\alpha =\left.\frac{\partial J}{\partial t}\right|_{t_0,a_0}
\equiv 1/t_0, \,\,\,
2\gamma=\left.\frac{\partial^2 J}{\partial a^2}\right|_{t_0,a_0}>0.
$$ 
As a result, $v_x$ takes a singularity as $\tau =t_0-t\to 0$:
\begin{equation}
\label{13}
v_x=\frac{v'_{0}(a_0)}{\alpha \tau + \gamma (a-a_0)^2}.
\end{equation}
Hence, one can see that the contribution from the singularity
to the integral $I_2$,
\begin{equation}
\label{13'}
I_2\approx\int \frac{v'^2_{0}(a_0)}{\alpha \tau + 
\gamma (a-a_0)^2}\,da \sim \frac{1}{\tau^{1/2}},
\end{equation}
diverges as $\tau\to 0$ and satisfies the inequality (\ref{10}).

\section{Multi-dimensional breaking}

In this section we  generalize the above
analysis to multi-dimensions.

In the multi-dimensional case, instead of $v_x$ it is convenient to introduce the
matrix $U$ with matrix elements
\begin{displaymath}
U_{ij}=\frac{\partial v_j}{\partial x_i} .
\end{displaymath}
The equations of motion for this matrix have a form
 analogous to (\ref{11}):
\begin{equation}
\label{14}
\frac{dU}{dt}=-U^2
\end{equation}
where
$$
\frac{d}{dt}=\frac{\partial}{\partial t}+({\bf v}\cdot\nabla).
$$
Our aim is now  to find the inequality corresponding to (\ref{8}).  Consider two
scalar characteristics of the matrix $U$: its trace, $\mbox{\rm tr}~ U\equiv \mbox{div}~{\bf v}$ and
determinant, $\det~ U$.  
From (\ref{14}) we derive the following
 equations for these two quantities:
\begin{equation}
\label{15}
\frac{d}{dt}\mbox {\rm tr}~ U=-\mbox {\rm tr}~(U^2),
\end{equation}
\begin{equation}
\label{16}
\frac{d}{dt}\det~ U=-\mbox {\rm tr}~U\cdot
\det~ U .
\end{equation}

Now we introduce  the positive definite integral,
\begin{displaymath}
I=\int (\det~U)^2~d{\bf r},
\end{displaymath}
which, in the 1D case, coincides with $I_2$.
Due to (\ref{16}), we have 
\begin{equation}
\label{17}
\frac{dI}{dt}=-\int~\mbox {\rm tr}~U ~(\det~U)^2d{\bf
r} .
\end{equation}
This equation generalizes   the equation (\ref{6}) to the multi-dimensional
case. 
The second derivative of $I$ will be given by the expression,
\begin{equation}
\label{18}
\frac{d^2I}{dt^2}=
\int \biggl [ (\mbox {\rm tr}~U)^2+
\mbox {\rm tr}(U^2) \biggr ] (\det U)^2~d{\bf r} .
\end{equation}
Applying the Cauchy-Bunyakovsky inequality to the r.h.s. of (\ref{16})
yields
\begin{equation}
\label{19}
\frac{dI}{dt}\le I^{1/2}\cdot\Biggl ( \int \mbox {\rm tr}(U^2)~
(\det U)^2 ~d{\bf r} \Biggr )^{1/2} .
\end{equation}
From (\ref{18}) we write the integral
\begin{equation}
\label{20}
\int (\mbox {\rm tr}~U)^2(\det~U)^2d{\bf r}=
\frac{d^2I}{dt^2}-\int \mbox {\rm tr}(U^2)
(\det~U)^2 d{\bf r}.
\end{equation}
Now, we shall estimate the  integral on the r.h.s. of this equation. 
We shall assume that all eigenvalues of the matrix $U$ are real. Such 
an assumption means that
the matrix $U$ is close to its symmetric part, $S=1/2(U+U^T)$, 
the so-called stress tensor (here $T$ denotes transpose).
In this case the antisymmetric part of the matrix $U$, the vorticity tensor,  
$\Omega=1/2(U-U^T)$ \footnote{The vorticity tensor $\Omega$ is connected with 
the vorticity ${\bf \omega}$
by the relation $\Omega_{ij}=\frac 12\epsilon_{ijk}\omega_k.$},
is small compared to $S$. 
In particular, if $\Omega=0$ the matrix
$U$ coincides with $S$, representing  the Hessian of the velocity potential $\Phi$:
$U_{ij}=S_{ij}=\partial^2 \Phi/\partial x_i\partial x_j $. 

Under this assumption the trace of the matrix $U$,
$\mbox {\rm tr}(U^2)=\sum\limits^D_{i=1}
\lambda^2_i$, where $\lambda_i$ are eigenvalues of the 
$U$ and $D$ dimension, becomes positive. In this case the following 
 relation between traces of $U$ and $U^2$ can be easily proven:
\begin{displaymath}
\sum\limits^D_{i=1}\lambda^2_i\ge \frac{1}{D}
\Biggl ( \sum\limits^D_{i=1}\lambda_i \Biggr )^2 .
\end{displaymath}
This inequality generates the following estimate between integrals: 
\begin{equation}
\label{21}
\int (
\det~U)^2 dr \int \mbox {\rm tr}~(U^2)
( \det~U)^2 dr \ge \frac{1}{D}
\Biggl ( \int \mbox {\rm tr}~U\cdot (\det U)^2~d{\bf r}
\Biggr )^2 \equiv
\frac{1}{D} \Biggl ( \frac{dI}{dt} \Biggr )^2~.
\end{equation}
Substitution (\ref{20}) and (\ref{21}) into (\ref{19}) gives the desired 
differential inequality,
\begin{equation}
\label{22}
I\frac{d^2I}{dt^2}-
\left ( 1+\frac{1}{D} \right )
\Biggl ( \frac{dI}{dt} \Biggr )^2\geq 0.
\end{equation}
Its solution is sought in  power form as before:
$I=x^{\alpha}$. Excluding terms proportional to $X^{\alpha -2}X^2_t$
we find that $\alpha=-D$ and 
$$
X=\frac{1}{I^{D}}.
$$
For $X$ this results in  
the same inequality as (\ref{9}): $X_{tt}<0$.
The criterion for attaining the origin $X=0$ will be also 
analogous:
\begin{displaymath}
X_t(0) <0 ~~\mbox {\rm or}~~ I_t(0) > 0~.
\end{displaymath}
Almost the same form will have the estimate for the collapse time
\begin{equation}
\label{t}
t_0 < \frac{DI(0)}{I_t(0)}=-\frac{1}{\langle\lambda(0)\rangle},
\end{equation}
where $\langle\lambda\rangle$ is a mean eigenvalue 
defined in accordance with (\ref{16}):
$$
\langle\lambda\rangle=\frac 1D\sum_i\bar\lambda_i =
\frac{1}{DI}\int~\mbox {\rm tr}~U ~(\det~U)^2d{\bf
r}.
$$
For arbitrary $D$,
instead of (\ref{10}), the following estimation appears for $I$:
\begin{equation}
\label{23}
I(t)\le \frac{I^{D+1}(0)}{(DI_t(0)(t_0-t))^D}.
\end{equation}

\section{Comparison with exact solution}

In order to compare the estimation (\ref{23}) with the exact dependence of $I$
we have to solve  equation (\ref{14}). This solution is
\begin{equation}
\label{sol}
U=U_0({\bf a})(1+U_0({\bf a})t)^{-1}.
\end{equation}
Here ${\bf a}$ is the initial coordinates of a fluid particle and 
$U_0({\bf a})$ is
the initial value of the matrix $U$. 
By introducing the projectors $P^{(k)}$ of the matrix
$U_0({\bf a})$ ($P^{(k)2}=P^{(k)}$ corresponding to each of the eigenvalues 
$\lambda_{0k}({\bf a})$),
this expression can 
be rewritten in the form of a spectral expansion:
\begin{equation}
\label{proj}
U=\sum \limits^D_{k=1}\frac{\lambda_{0k}}{1+\lambda_{0k}t} P^{(k)}.
\end{equation}
The projector $P^{(k)}$, being 
a matrix function of ${\bf a}$, is expressed through the eigenvectors
for the direct ($U_0({\bf a})\psi=\lambda_{0} \psi$) and conjugated ($\phi U_0({\bf a})
=\phi\lambda_{0}$) spectral problems for 
the matrix $U_0({\bf a})$:
$$
P^{(k)}_{ij}=\psi^{(k)}_i\phi^{(k)}_j.
$$
where the vectors $\psi^{(n)}$ and $\phi^{(m)}$ with different $n$ and $m$
are mutually orthogonal:
$$
\psi^{(m)}_i\phi^{(n)}_i=\delta_{mn}.
$$
Hence, the determinant of the matrix $U$ is defined by the product,
$$
\det~U= \prod\limits^D_{k=1}\frac{\lambda_{0k}}{1+\lambda_{0k}t}.
$$
From (\ref{proj}) it follows also that singularity in $U$ first time appears at $t=t_0$,
defined from the condition  \cite{5,frisch}:
\begin{equation}
\label{min1}
t_0= \min_{k,a}[-1/\lambda_{0k}({\bf a})],
\end{equation}
(compare this with (\ref{t})).

From (\ref{proj}), one can see  that 
near the singular point  only one term in the sum (\ref{proj}) survives,
\begin{equation}
\label{col}
U\approx -\frac{P^{(n)}}{\tau + \gamma_{\alpha\beta}
 \Delta a_{\alpha} \Delta a_{\beta}},
\end{equation}
where the projector $P^{(n)}$ is  evaluated at the point ${\bf a=a_0}$ and
 $k=n$, 
corresponding to the minimum
(\ref{min1}), $\tau=t_0-t$, $\Delta{\bf a=a-a_0}$, and
$$
2\gamma_{\alpha\beta}=-\left.\frac{\partial^2 \lambda_{0n}^{-1}}{\partial a_{\alpha}
\partial a_{\beta}} \right|_{a=a_0}
$$ 
is a positive definite matrix. 

The remarkable formula (\ref{col}) demonstrates that i) the matrix $U$
tends to the degenerate one as $t\to t_0$ and ii)
both parts of the matrix $U$ in this limit, i.e. the stress 
tensor $S$ and the vorticity 
tensor $\Omega$, become simultaneously infinite
(compare with \cite{3}). 
It is interesting to note that
at near singular time the ratio between both parts is fixed  
 and governs by two relations following from the definition of the 
 projector $P$:
$$
P_S=P_S^2+P_A^2,\,\,\, P_A=P_SP_A+P_AP_S
$$
where $P_S$ and $P_A$ are respectively symmetric ("potential") 
and antisymmetric 
(vortical) parts of the projector $P$. In particular, the second relation 
provides
the collapsing solution for the equation for vorticity
$$
\frac{\partial \bf \omega}{\partial t}=\mbox{curl}~[{\bf v}\times{\bf \omega}]
$$
which has the same form for both compressible and incompressible cases. 
It is also interesting to note
that in the sense of the criterion (\ref{1}), the collapsing solution 
(\ref{col}) 
represents the marginal solution.

The asymptotic solution (\ref{col}), far from the collapsing point 
should be matched with a "regular" solution. The corresponding
matching scale can be estimated
as $l_0\approx \gamma^{2/3}.$
This scale $l_0$ alone can be taken as the size  of the collapsing region 
for (\ref{col}). 
This remark now allows one to calculate the  contribution from
 the breaking area to the integral $I$.

Substituting (\ref{sol}) into $I$ and using a change of variables, from
${\bf r }$ to ${\bf a}$, one can get the expression for this contribution,
\begin{equation}
\label{int}
I=C \int_V \frac{d^D a}{\tau + \gamma_{\alpha\beta}
 a_{\alpha} a_{\beta}},
\end{equation}
where the constant is
$$
C=\left.\lambda_{0n}\det U_0\prod\limits_{k\neq n}\frac{\lambda_{0k}}
{1+\lambda_{0k}t_0}\right|_{a=a_0}.
$$
The integral is taken over the spherical volume $V$ with coordinate center at
 ${\bf a=a_0}$ and size  $\sim {l_0}$.
Introducing a self-similar variable ${\bf \xi}={\bf a}\tau^{-1/2}$, 
one can see
that the contribution depends significantly  on the dimension $D$. 
At $D=1$ this integral
behaves like $\tau^{-1/2}$ in full correspondence with (\ref{13'}). 
In this case, 
the integral over $\xi$ is convergent at large  $\xi$ and is not sensitive to 
the cut-off size $l_0$. In the two-dimensional geometry, however, the integral 
(\ref{int}) has a power dependence on $\tau$, but a logarithmic
dependence on $l_0$ arising from integration on $\xi$:  
$$
I\sim \log\frac{l_0}{\tau^{1/2}}
$$ 
that  satisfies the inequality (\ref{23}).

In the three-dimensional case, the integral (\ref{int}) diverges at 
large scales as the first power 
of $\xi$, becomes  proportional to the size of collapsing area 
$l_0$ as $\tau\to 0$:
$$
I\sim l_0.
$$
This result for $D=3$ formally contradicts to 
the blow-up sufficient condition 
found above. This contradiction indicates only that 
in the three-dimensional case  the blow-up
of the integral $I$  has no universal behavior near the 
singular time which should be
expected following to 
the universal asymptotics (\ref{col}).

\section{Concluding remarks}

Thus, the initial condition,
\begin{displaymath}
\frac{dI}{dt}(0)>0,
\end{displaymath}
represents a sufficient integral criterion for the collapse if
 the vorticity matrix $U$ is small in comparison to the stress tensor. 
 Under this
condition, the integral $I=\int (\det U)^2d{\bf r}$
becomes infinite in a finite time. 

In  turn, divergence of $I$ means 
that a solution can not be continued 
in the corresponding functional  space. 
In the $1D$ case this is the Sobolev space $H^1({\cal R})$. For
$D=2$, from the inequality,
\begin{displaymath}
I\le \int |U|^{4}d^2{ r}\,\,\,\, (|U|^2\equiv U^2_{ik}),
\end{displaymath}
together with the embedding Sobolev
inequality,
\begin{displaymath}
\Biggl ( \int |U|^4d^2{\bf r} \Biggr )^{1/4}\le C\| U \|_{H^1({\cal R}^2)}
\end{displaymath}
it follows that $\| U \|_{H^1({\cal R}^2)}\to\infty$ as $I\to\infty$. 
In terms of the velocity, this means that the solution is not continued
in the Sobolev space $H^2({\cal R}^2)$.

 In the $3D$ case it is possible to write the following set of inequalities,
\begin{displaymath}
I^{1/6}\le \left(\int |U|^6d^3{\bf r}\right)^{1/6}\le C\| U\|_{H^1({\cal R}^3)},
\end{displaymath}
where the second inequality represents  the partial case 
of the Sobolev embedding inequality  \cite{8}.

Hence, one can see that
 $I\to\infty$ is equivalent to the divergence of the norm (\ref{norm})
 for the Sobolev space  $H^2({\cal R}^3)$. Thus, the requirements 
for strong solutions in the hydrodynamic model (\ref{2}), (\ref{3}) 
are different from those for the 3D Euler equation for incompressible fluids.
At the moment it is hard to say whether the results presented in this paper
contradict  to the theorem \cite{2}  (see also \cite{7}) or not.  In any case,
it is a very interesting question. It should be added 
that  collapse in incompressible fluids might happen through  breaking of 
vortex lines (there are some 
arguments both analytical and numerical \cite{6,ZKP} 
in a favor of such point 
of view). In this case  
 the corresponding norms would also blow up  for the same Sobolev space
$H^2({\cal R}^3)$.

\section{Acknowledgments}
The author would like to thank to P.L. Sulem for helpful discussion
and to R. Grauer who paid the author' attention to the paper \cite{3}. 
The author is grateful to the Observatory of Nice, where this work was
initiated within the Landau-CNRS agreement. This work was 
supported by RFBR (grant no.~00-01-00929) 
and by INTAS (grant no. 00-00292).

\end{document}